\documentclass[journal,10pt]{IEEEtran}

\usepackage{pifont,bm,multicol,amsfonts,amsmath,color,amssymb,graphicx, epsfig,cite,psfrag,subfigure,algorithm,enumerate,stfloats,algorithm,algorithmic,epstopdf}

\newtheorem{theorem}{\underline{Theorem}}

\IEEEoverridecommandlockouts

\begin{document}

%
\title{Base Station Selection for Massive MIMO Networks with Two-stage Precoding}

\author{
	Jianpeng Ma, { \emph{Student Member, IEEE,}} Shun Zhang, { \emph{Member, IEEE,}} Hongyan Li, \emph{Member, IEEE,} \\Nan Zhao, \emph{Senior Member, IEEE}
and Victor C.M. Leung, \emph{Fellow, IEEE}
\thanks{J. Ma, S. Zhang, and H. Li  are with the State Key Laboratory of Integrated Services Networks,
	Xidian University, Xi'an 710071, P. R. China. $\big($Email:
	jpmaxdu@gmail.com; zhangshunsdu@gmail.com; hyli@mail.xidian.edu.cn$\big)$.}
\thanks{N. Zhao is with the School of Inform. and Commun. Eng., Dalian University of Technology, Dalian, Liaoning, P. R. China (Email: zhaonan@dlut.edu.cn).}
\thanks{V.C.M. Leung is with the Department of Electrical and Computer Engineering,
	the University of British Columbia, Vancouver, BC, V6T 1Z4, Canada (Email: vleung@ece.ubc.ca).}
}


\maketitle \thispagestyle{empty} \vspace{-10mm}


%


\maketitle

\begin{abstract}
	The two-stage precoding
	has been proposed  to reduce the overhead of  {both the}
channel training  and  {the} channel state information (CSI) feedback for  the massive  multiple-input multiple-output (MIMO) system.
	But the overlap  of    {the}
angle-spreading-ranges (ASR) for different user clusters may seriously degrade the performance of   {the} two-stage precoding.
In this letter, we propose one ASR overlap mitigating scheme through   {the}
base station (BS) selection. Firstly, the BS selection is formulated as a sum  signal-to-interference-plus-noise ratio (SINR) maximization problem. Then, the problem is solved by a  {low-complex} algorithm through maximizing signal-to-leakage-plus-noise ratio (SLNR). In addition, we propose one low-overhead algorithm with the lower bound  {on} the average SLNR as the objective function.
Finally, we demonstrate the	efficacy of the proposed schemes through  {the} numerical simulations.
\end{abstract}

\maketitle \thispagestyle{empty} \vspace{-1mm}


\maketitle \thispagestyle{empty} \vspace{-3mm}

%
\IEEEpeerreviewmaketitle

\section{Introduction}
The massive multiple-input multiple-output (MIMO) system has been considered as a promising technology to meet the capacity demand in  the  next generation wireless cellular networks \cite{Massive_in_5G_1}.
In  {the} frequency-division duplex  (FDD) system where the uplink-downlink reciprocity does not exist,   {the} CSI at the BS side should be obtained through {the} downlink training  and  {the}
CSI feedback, which will lead to  {the} unacceptable  {overheads} \cite{training_fdd}. {To overcome this bottleneck},
the joint spatial-division and multiplexing  (JSDM)
was  {recently proposed as one
two-stage precoding scheme}\cite{JSDM}.

Several researchers followed the  two-stage precoding and developed  different  prebeamforming  {methods}\cite{JSDM,TQF,iterative,beam_division}.
In \cite{JSDM}, Adhikary \emph{et al.} proposed one block diagonalization (BD) algorithm, whose main idea is  projecting the channel eigenspace of the desired cluster into  that for all the other clusters.   In \cite{TQF}, an iterative prebeamforming algorithm was designed to maximize the signal-to-leakage-plus-noise ratio (SLNR).  In \cite{iterative}, Chen and Lau developed a  {low-complex} online  iterative algorithm to track the  prebeamformer. In  \cite{beam_division}, Sun \emph{et al.} proposed the beam division multiplex access (BDMA) scheme to complete the optimal downlink transmission with only   statistical CSI.

\begin{figure}[!t]	
	\centering{\includegraphics[width=80mm]{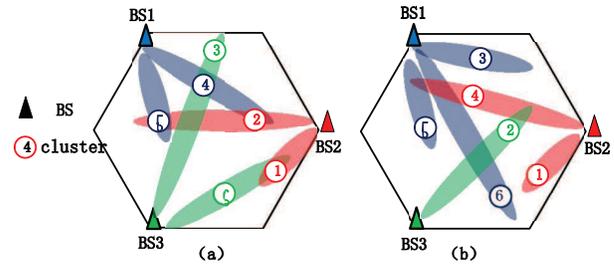}}
	\caption{  An downlink massive MIMO network with $L$ BSs, where clusters are randomly distributed. (a) One BS selection solution with serious ASR overlap. (b) Another    BS selection solution with slight ASR overlap.  }
	\label{fig:system}
\end{figure}

As mentioned in \cite{JSDM,TQF,iterative,beam_division}, when the angle-spreading-ranges (ASRs) of
the scattering rays for different clusters do not overlap,  {the}
orthogonal transmission can be
achieved with the two-stage precoding. However, in practice,   {the}
ASRs for some clusters may overlap,
which would degrade the performance of  {the}
two-stage precoding, especially in the high dense coverage scenario.
 {Furthermore, the} authors in \cite{JSDM,TQF,iterative,beam_division}  {only examined} the single BS scenario and  {scheduled the} clusters without (or with slight) ASR for transmission.
\textcolor{blue}{ In this letter,
we consider one  {typical} multiple BSs coverage scenario as shown in
	 \figurename{ \ref{fig:system}},}  {where  three  \textcolor{blue}{BSs}  with sector antennas of 120  degrees opening} are equipped at  {the} cell corners \cite{sector}.
In such scenario, the two user clusters with serious ASR overlap at  one BS  may have no or slight ASR overlap at the others.  For example, as shown in \figurename{ \ref{fig:system}},  the clusters  2 and 4 have serious ASR overlap seen from BS1,
but have no overlap  when they are connected  to BS2 or BS3.
Hence, the serious ASR overlap may be mitigated through  {the} BS selection, which can be formulated as a combinational optimization targeting at
maximizing the sum SINR for all users. \textcolor{blue}{ In \cite{Association},
	the authors  {prove}
	that if the BS selection problems  {are}
	considered together with the power control, then the optimal solution can be obtained.
	However, when the power is fixed, the optimal problem is  NP-hard.} To solve the optimization problem with low complexity, we develop one  SLNR-based BS selection algorithm. It is proved that the proposed low complexity algorithm can achieve the optimal   BS selection solution with the first   category prebeamformers.
Furthermore, we develop  one low-overhead suboptimal algorithm, whose objective function is
the lower bound of the average SLNR.
Finally, numerical simulations are presented to validate the proposed algorithms.

\section{System Model and Preliminaries}

\subsection{System Configuration and Channel Model}
As shown in \figurename{ \ref{fig:system}}, we consider a  downlink massive MIMO network with $L$ BSs,  where each BS  employs an  uniform linear array (ULA) with $N\gg 1$ antennas.  It is assumed that
 the $K$ single-antenna users can be  partitioned into $C$ clusters, where the users in the same cluster are almost co-located.  We denote the set of all clusters as $\mathcal C = \{1,2,\cdots, C\}$. As a result, the number $K_c$ of the users in the cluster $c$  satisfies  $\sum\limits_{c=1}^{C} K_c =K$.
 The $1 \times N$ downlink channel vector from the BS $l$ to the user $k$  in
 the cluster $c$   is
 $\{\mathbf h_{c,k}^l \}^T$ and is assumed as block fading. We further define the $K_c \times N$ downlink channel matrix between the BS $l$ and  the  cluster $c$ as $\mathbf H_c^l  = [ \mathbf h_{c,1}^{l }  ,\mathbf h_{c,2}^{l },  ,\cdots,\mathbf h_{c,{K_c}}^{l }  ]^T$.

 Similar to the works  \cite{JSDM,iterative,TQF,beam_division},  the classical  ``one-ring"
 \footnote{\textcolor{blue}{For the sake of clarity, we only consider the  ``one-ring'' scenario in Fig. 1.   Nevertheless, the proposed  scheme can be  also applicable for the practical scenarios with multiple scattering rings.}}
  model \cite{onering} is applied to describe the channels.
 It can be readily checked that all the users in the same cluster share the same one-ring model parameters, which means that the users in the same cluster have the same $N \times N$ channel covariance matrix $\mathbf R_c^l$ at BS $l$, i.e.,
$
 \mathbb E\Big\{ \mathbf h_{c,k}^l \big[{\mathbf h_{c,k}^l}\big]^H\Big\} =  \mathbf R_c^l,  k= 1,2,\cdots K_c.$
  Furthermore, we can express $\mathbf R_c^l$ as
 \begin{equation}\label{def-R}
 \left[\mathbf{R}_{c}^l\right]_{p,q}  =
 \frac{1}{2\Delta _{c}^l} \int _{\theta_{c}^l -\Delta_{c}^l }^{\theta_{c}^l +\Delta_{c}^l}
 e^{\frac{-2i\pi (p-q)  \sin(\alpha) \tau}{\lambda}} d\alpha,
 \end{equation}
 where $\theta_{c}^l$ is the azimuth angle of the scatters ring,  $[\theta_{c}^l - \Delta_{c}^l, \theta_{c}^l + \Delta_{c}^l ]$ is  the  ASR of the cluster $c$ at the BS $l$,  $ \tau$ is the antenna element spacing, and $\lambda$ is the carrier wavelength.

 With eigen-decomposition, we have $  \mathbf{R}_{c}^l  = \mathbf{E}_{c}^l    {\bm{{\Lambda}}_{c}^l    }      \big[\mathbf{E}_{c}^l  \big]^H$, where  the  diagonal matrix {${\bm{{\Lambda}}_{c}^l }$}
 is the nonzero eigenvalue matrix, {${\mathbf{{E}}_{c}^l  }$} contains the eigenvectors corresponding to the nonzero eigenvalues, and  the rank of $\mathbf{R}_{c}^l  $,  \emph{i.e., $r_{c}^l  $}, is  much smaller than  $N$.
 Resorting to the Karhunen-Loeve representation, we can {write}
 the channel vector
 $\mathbf{h}_{c,k}^l$ as~\cite{JSDM}
 \begin{align} \label{massive MIMO channel}
 \mathbf{h}_{c,k}^l=
 {\mathbf{E}_{c}^l}{\left\{\bm{{\Lambda}}_{c}^l\right\} }^{\frac{1}{2}}
 \mathbf w_{c,k}^l,
 \end{align}
 where  the entries of the $r_{c}^l \!\times\! \!1\!\!$ vector
 $\mathbf w_{c,k}^l\!$ are i.i.d. complex Gaussian distributed with zero mean and unit variance.

  Let us represent the  cluster set accessed to the BS $l$ as $\mathcal C_l$, which satisfies $\cup_{l=1}^L \mathcal C_l =\mathcal C$. Then,
 the received signal at the cluster $c$ accessed to the BS $l$ is
\begin{align}\label{recv_interference}
\mathbf y_{c\in \mathcal C_l} =& \mathbf H_c^l   \mathbf P_c^l    \mathbf d_c^l
+ \sum_{c' \in \mathcal C_l,c'\neq c} \mathbf H_{c}^l    \mathbf P_{c'}^{l}    \mathbf d_{c'}^{l}  \notag \\
&~~~~~~~~~~~	+\sum_{l'\neq l}^{L}  ~\sum_{c' \in \mathcal C_{l'}} \mathbf H_{c}^{l'}    \mathbf P_{c'}^{l'}   \mathbf d_{c'}^{l'}   + \mathbf n_c,
\end{align}
where $\mathbf P_{c}^l$  is the $N \times S_c$  precoding matrix for the cluster $c$ at the BS $l$,
$\mathbf d_{c}^l$ with $\mathbb{E}\{\mathbf d_{c}^l (\mathbf d_{c}^l)^H\} = \mathbf I$ is the
$S_c^l \times 1$ data vector from BS $l$ to the cluster $c$, and the $K_c\times 1$ vector $\mathbf{n}_{c}\! \sim \!\mathcal{CN}(\mathbf 0, \sigma_n^2 \mathbf{I}_{ K_c})$ denotes the additive white Gaussian noise.
\subsection{Two-stage Precoding}

As in  \cite{JSDM,iterative,TQF,beam_division},  the  two-stage precoding can be written as
\begin{align}
\mathbf P_{c}^l=\mathbf B_{c}^l \mathbf V_{c}^l,
\end{align}
 where the $N\times M_c^l $ prebeamforming matrix $\mathbf B_c^l$ only depends on  the
 channel covariance  matrices, and is used to eliminate the inter-cluster interferences;
the $M_c^l\times S_c^l$ inner precoder   $\mathbf V_{c}^l = [ \mathbf v_{c,1}^l, \mathbf v_{c,2}^l, \cdots, \mathbf v_{c,K_c}^l]$, dealing with the intra-cluster interferences, can be designed with the $K_{c} \times M_c^l$ equivalent channel matrix $\overline{\mathbf{H}}_c^l= {\mathbf{H}}_c^l \mathbf{B}_c^l$;
\textcolor{blue}{ $M_c^l$ } is the dimension of $\overline{\mathbf{H}}_c^l$ seen by the inner precoder, {and} satisfies $S^l_c \leq M_c^l \leq r_c^l$.
It can be found that $\overline{\mathbf{H}}_c^l$ possesses a much smaller number of unknown parameters than the original channel matrix $\mathbf{H}_{c}^l $.

 The prebeamforming matrix designing methods can be divided into two categories
according to {the fact}
whether the inter-cluster interference is completely eliminated:

 {\subsubsection{\textcolor{blue}{ The first category of the prebeamformers}} the inter-cluster interference is eliminated  completely by designing $\mathbf B_c^l $ to satisfy the following constraint,
\begin{align}
(\mathbf E_{c^{\prime}}^l)^H \mathbf B_c^l = 0, \forall c^{\prime} \neq c.
\end{align}
The BD method in \cite{JSDM},  BDMA method in \cite{beam_division}, and DFT-based method in \cite{gao_E} fall into this category.}

\subsubsection{\textcolor{blue}{The second category of the prebeamformers}} 
the inter-cluster interference is not completely eliminated, which can be  expressed mathematically as
\begin{align}
(\mathbf E_{c^{\prime}}^l)^H \mathbf B_c^l \approx 0, \forall c^{\prime} \neq c.
\end{align}
The approximate BD method in \cite{JSDM} and methods in \cite{TQF} and \cite{iterative} belong to this category.

\section{Proposed User Scheduling Scheme}

\subsection{Problem Formulation}

With prebeamforming, the inter-cluster interference is completely (or almost) eliminated. Without loss of generality, we assume that the intra-cluster interference is completely eliminated by the zero-forcing  inner precoder $\mathbf V_c^l$. Then, the SINR of  the user $k$ in the cluster $c \in \mathcal C_l$ can be derived from \eqref{recv_interference} as follow,
\begin{align}\label{SINR}
 &\text{SINR}_{c,k}^l= \frac{|(\mathbf h_{c,k}^l)^T \mathbf B_{c}^l \mathbf v_{c,k}^l|^2}{\gamma_{c,k}^l + \sigma_{n}^2},
 \end{align}
 {where}
\begin{align}
 \gamma_{c,k}^l =&\sum_{c' \in \mathcal C_l,c'\neq c} ~ \sum_{k'=1}^{K_{c'}}|( \mathbf h_{c,k}^l)^T    \mathbf B_{c'}^l     \mathbf v_{c',k'}^l  |^2 \notag\\
&+ \sum_{l'=1,l'\neq l}^{L}~\sum_{c' \in \mathcal C_l}~ \sum_{k'=1}^{K_{c'}}|( \mathbf h_{c,k}^{l'})^T    \mathbf B_{c'}^{l'}    \mathbf v_{c',k'}^{l'} |^2
\end{align}
is  the power of the inter-cluster interference. Specifically, we get $\gamma_{c,k}^l  = 0$, if
the first category prebeamforming is applied. Different from the classical user scheduling schemes \cite{JSDM,TQF,iterative,beam_division},
  we mitigate the inter-cluster interference through   BS selection for each user cluster.
For clear illustration, we present two examples in Fig. 1.
  With respect to one specific   clusters distribution scenario,
  the BS selection scheme in Fig. 1-b suffers slighter ASR overlap  than that in Fig. 1-a.
Given $C$ clusters and $L$ BSs,
  the total number of the BS selection strategies is $L^C$.  { Since the ASR overlap  decreases the desired signal power and increases the inter-cluster interference power,
  our task  is to find the best strategy from all the possible ones to maximize the sum SINR}, and the corresponding optimization problem can be formulated as
\begin{align}
(\text P1) ~~~&\max_{\mathcal C_1, \mathcal C_l, \cdots, \mathcal C_L } \sum_{l=1}^{L}\sum_{c \in \mathcal C_l} \sum_{k=1}^{K_c} \text{SINR}_{c,k}^l\\\notag
&~~~~~\text{s.t.} ~~ \cup_{l=1}^{L} \mathcal C_l =\mathcal C.
\end{align}

The optimal problem (P1) is NP-hard,
since the SINR of one specific cluster is closely related to the other user  {clusters'} BS selection strategy. Theoretically, it can be solved by the exhaustive search with the complexity $\mathcal O(L^C)$, which is unacceptable for large $C$ and $L$. To reduce the complexity,  we adopt  SLNR  as another optimization criterion.
%

%

\subsection{ SLNR-Based BS Selection Algorithm with Low Complexity}
 The SLNR is defined as the ratio between the signal power to the desired receiver and total interference power to the undesired receivers, which is {commonly used if} the maximizing SINR problem is difficult to solve.
The SLNR of the user $k$ in the cluster $c$ can be written as
\begin{align}\label{SLNR}
\text{SLNR}_{c,k}^l =& \frac{|(\mathbf h_{c,k}^l)^T\mathbf B_{c}^l \mathbf v_{c,k}^l|^2}{\zeta_{c,k}^l + \sigma_{n}^2},\\
\text{where~~~~~~~~~~ }
\zeta_{c,k}^l =& \sum_{c' \in \mathcal C,c'\neq c} ~ \sum_{k'=1}^{K_{c'}}| (\mathbf h_{c',k'}^l)^T   \mathbf B_{c}^l
    \mathbf v_{c,k}^l  |^2 \notag~~~~~~~~~~~~~~~~
\end{align}
is the leaked interference power. Specifically, $\zeta_{c,k}^l  = 0$, for the first category prebeamforming. Then, the BS selection problem can be reformulated as
\begin{align}
(\text P2) ~~~&\max_{\mathcal C_1, \mathcal C_l, \cdots, \mathcal C_L } \sum_{l=1}^{L}\sum_{c \in \mathcal C_l} \sum_{k=1}^{K_c} \text{SLNR}_{c,k}^l\\\notag
&~~~~~\text{s.t.} ~~ \cup_{l=1}^{L} \mathcal C_l =\mathcal C.
\end{align}
The main difference between (P1) and (P2)
can be summarized as follows: the SINR of one specific user depends on  all other clusters' BS selection strategies, while the SLNR is only related to its own selection decision.
Thus, to achieve the optimal solution of (P2), we can separately implement the BS selection operation of each user cluster
to maximize its own sum-SLNR, which is presented in \textbf{Algorithm \ref{Algorithm1}}.
\begin{theorem} \label{prop2}
	 When the prebeamforming methods in the first category is used, the SLNR-based BS selection algorithm achieves the optimal solution for the  SINR maximization problem (P1).
\end{theorem}
\begin{IEEEproof}	If the first category methods is applied, both $\gamma^l_{c,k} $  and  $\zeta^l_{c,k} $ are zero.
Therefore, the SLNR of a user is same with its SINR, which indicates that  the  problems (P2)  and (P1) are equivalent and have
 the  same optimal solution.
\end{IEEEproof}

 \textcolor{blue}{In    Algorithm 1, one cluster is assigned to a BS in  {the steps } (4--7)  {of each iteration}. Therefore,  only $C$ iterations are required to complete BS selection for all $C$ clusters. Since  {only} one BS is selected from $L$  {ones} in each iteration, the complexity of   the  Algorithm 1 is $\mathcal O(LC)$,} which is much smaller than that of  the exhaustive search. Take $C=25,L=3$ as a example, $\frac{LC}{L^C}= 8.85\times 10^{-11} $.

\begin{algorithm}[!t]
	\caption{SLNR-Based  BS Selection Algorithm  }
	\label{Algorithm1}
	\begin{algorithmic}[1]
		\REQUIRE {   $\{(\mathbf h_{c,k}^l)^H \mathbf B_c^l\}$,$\{\mathbf V_c^l\}$  }.
		\ENSURE{The clusters set that select each BS : $\mathcal C_1, \mathcal C_2, \cdots, \mathcal C_L$}.
		
		\STATE{	\textbf{Initialization:} }
		\STATE	{	$\mathcal{C}_l = \emptyset, ~\forall l=1,2,\cdots, L.$}
		\STATE \textcolor{blue}{$c=1$.}
		\REPEAT
		\STATE	{$ l^*_c = \arg \max_l \sum_{k=1}^{K_c} \text{SLNR}^l_{c,k}  $, $ \mathcal C_{ l^*_c }   =   \mathcal C_{ l^*_c } \cup \{ c\} $.	}
		\STATE \textcolor{blue}{ $c= c+1$.}
		\UNTIL{$ c > C$}		
		\RETURN{$\mathcal C_1, \mathcal C_2, \cdots, \mathcal C_L$.}
	\end{algorithmic}
\end{algorithm}
\subsection{ {LASLNR-Based BS Selection Algorithm with Low Overhead} }
The previous subsection  presents a  {low-complex} BS selection algorithm. However,  it
requires  the instant CSI of the  equivalent effective channels  between all  the  users and  the BSs. Acquiring such a large amount of the equivalent channels would consume too much channel bandwidth.
Therefore, we intend to give a low overhead BS selection method based on the average SLNR, which would   only need the prebeamforming matrices $\{\mathbf{B}_c^l\}$ and channel covariance matrices $\{\mathbf R_c^l\}$. Unfortunately, it is very difficult to derive the closed-form expressions for the average SLNR. Thus, we derive a close-form lower bound on the
average SLNR (LASLNR)  as  {follows:}

 \textcolor{blue}{
\begin{align}\label{SLNR_lower_bound}
&\mathbb E \{ \text{SLNR}_{c,k}^l \}\notag \\
&\stackrel{(a)}{\geq} \mathbb E \Bigg \{\frac{|(\mathbf h_{c,k}^l)^T\mathbf B_{c}^l \mathbf v_{c,k}^l|^2}{
	P_t \sum_{c' \in \mathcal C,c'\neq c} ~ \sum_{k'=1}^{K_{c'}}\| (\mathbf h_{c',k'}^l)^T   \mathbf B_{c}^l
	  \|^2 \notag  + \sigma_{n}^2} \Bigg\}\notag\\
	&\stackrel{(b)}{\geq}  \frac{ \mathbb E  \big \{  |(\mathbf h_{c,k}^l)^T\mathbf B_{c}^l \mathbf v_{c,k}^l|^2 \big  \}}{
		 \mathbb E \Big \{  P_t \sum_{c' \in \mathcal C,c'\neq c} ~ \sum_{k'=1}^{K_{c'}}\| (\mathbf h_{c',k'}^l)^T   \mathbf B_{c}^l
		\|^2 \notag  + \sigma_{n}^2 \Big\}}\notag\\
	&\stackrel{(c)}{\geq}  \frac{ \text{tr}\Big\{(\mathbf B_{c}^l)^H \mathbf R_{c}^l \mathbf B_{c}^l \!\Big\} \!\!- \!\!(K_c-1)  \lambda_c^l}{
		 \sum_{c' \in \mathcal C,c'\neq c} ~ \sum_{k'=1}^{K_{c'}}\text{tr}\Big( (\mathbf B_{c}^l)^H  \mathbf R_{c'}^l  \mathbf B_{c}^l \Big	)
		  + \sigma_{n}^2/P_t}\notag\\
	&= \text{LASLNR}_{c,k}^l,
\end{align}}
where $\lambda_c^l$ is the largest eigenvalue of $\mathbf R^l_c$ and
 $P_t = \|\mathbf v_{c,k}^l\|^2$  is the transmitting power for each user. The derivation of (a) utilizes the matrix property equations $\|\mathbf A \mathbf B\| \leq \|\mathbf A\| \| \mathbf B\|$ and $\|\mathbf v^l_{c,k} \| =P_t$; (b) results from Jensen¡¯s inequality and the fact that $\mathbf h_{c,k}^l  $ and $\mathbf h_{c',k'}^l  $  are independent;  (c)  follows from Theorem  1 of \cite{TQF}.

Through replacing the step 4 in \textbf{ Algorithm \ref{Algorithm1}} with
\begin{align}
l^*_c&= \arg \max_l \sum_{k=1}^{K_c} \text{LASLNR}^l_{c,k},
\end{align}
we can get the   LASLNR-based low-overhead BS selection  algorithm.

\section{Numerical Results}

In this section, we evaluate the performance of our proposed algorithms through numerical simulations. A typical  single-cell with radius 1km is considered, where $L = 3$ BSs are equipped at three cell corners. Each BS is equipped with a ULA of $N = 128$ antennas.
\textcolor{blue}{The number of {the} users in each cluster is set as 3.}  $C$ clusters are uniformly randomly distributed in the cell. The carrier frequency is $2$GHz. The massive MIMO channel is generated by \eqref{def-R} and \eqref{massive MIMO channel}.
The variance of the noise is $\sigma_n^2 =1$.
The  BD and the approximated BD (ABD) methods  in \cite{JSDM} are used as the representatives for the first and second categories of the prebeamforming methods, respectively.

\begin{figure}[!t]
	\centering
	\includegraphics[width=70mm]{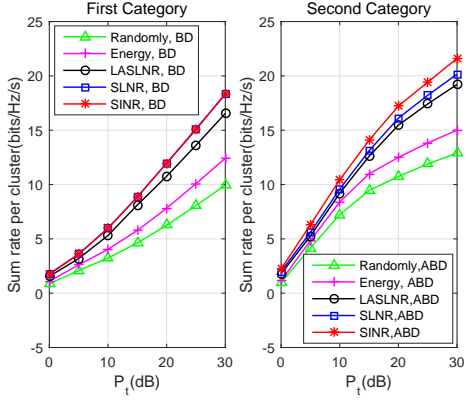}
	\caption{ Sum-rate per cluster  versus \textcolor{blue}{$P_t$} with $C =8, L =3$.}
	\label{fig:sum_rate_SNR}
\end{figure}


\figurename{ \ref{fig:sum_rate_SNR}}   presents the curves of the sum-rate per cluster  {versus} \textcolor{blue}{$P_t$}, where the first and second categories  of the  prebeamforming methods are used, respectively. The proposed  SLNR-based and LASLNR-based suboptimal algorithms are compared with the following methods: the exhaustive search for (P1), \textcolor{blue}{ the scheme based on  the largest energy of the
channel coefficient matrix, and the random BS selection scheme.
As observed from both figures, the proposed algorithms significantly outperform the  randomly BS selection scheme and  and the largest energy scheme.} The LASLNR-based suboptimal  algorithm almost achieves the same performance as the SLNR-based one. As  mentioned in Theorem 1,   the SLNR-based algorithm achieves the optimal solution  {for}  (P1) when the first category prebeamforming method is used, which is validated in \figurename{ \ref{fig:sum_rate_SNR}}. Moreover, it is shown in  \figurename{ \ref{fig:sum_rate_SNR}} that when the second category prebeamforming method is used the performance gap between SLNR-based algorithm  and the optimal solution is also small.

\begin{figure}[!t]
	\centering
	\includegraphics[width=70mm]{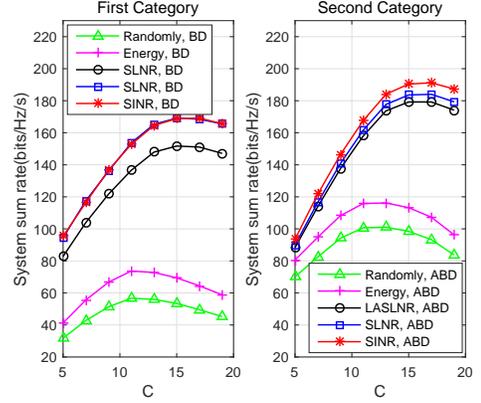}
	\caption{ System sum-rate  versus $C$  with   \textcolor{blue}{$P_t$} = 20dB and $ L =3$.}
	\label{fig:sum_rate_C}
\end{figure}


In \figurename{\ref{fig:sum_rate_C}}  the impact of the  number of clusters $C$  on the system sum-rate is demonstrated, where the first and second categories prebeamforming methods are used respectively. As we can see from the both figures, with the increase of $C$, the system sum-rate of all solutions increase at first and then decrease. \textcolor {blue} {The reason behind this is when $C$ is small, the ASR overlap is slight.  {Under this scenario}, increasing $C$ will result in more served users and higher system sum-rate. As $C$  {continues} to increase, the cell  {becomes crowded}  and the ASR overlap  {becomes} serious, which will degrade the system sum-rate.}  We can also see that the performance gaps between the proposed algorithms and the randomly selection solution become larger as  $C$ increases. These observations indicate that the proposed scheme is effective to mitigate the ASR overlap, especially when  the network is crowded.

\section{Conclusions}
In this letter, we have proposed a BS selection  scheme to  mitigate ASR overlap  for massive MIMO system with two-stage precoding.
The  BS selection problem was formulated as a combinatorial  optimization problem targeting  at maximizing the sum SINR. Then, we transformed the problem into a SLNR maximization optimization, which has low computational complexity. In order to further reduce the signaling overhead of SLNR maximization problem, a suboptimal solution was obtained by using the LASLNR as the objective function. The simulation results demonstrated that the proposed algorithms  significantly improve the sum rate performance and  the LASLNR-based algorithm achieves almost the same performance as the SLNR-based algorithm.






\begin{thebibliography}{1}

	
	\bibitem{Massive_in_5G_1}
 F.~Rusek, D.~Persson, B.~K. Lau, E.~G. Larsson, T.~L. Marzetta, O.~Edfors, and
 F.~Tufvesson, ``Scaling up {MIMO}: Opportunities and challenges with very
 large arrays,'' \emph{{IEEE} Signal Process. Mag.}, vol.~30, no.~1, pp.
 40--60, Jan. 2013.


	
	\bibitem{training_fdd}
 S.~Noh, M.~D. Zoltowski, and D.~J. Love, ``Training sequence design for
 feedback assisted hybrid beamforming in massive {MIMO} systems,''
 \emph{{IEEE} Trans. Commun.}, vol.~64, no.~1, pp. 187--200, Jan. 2016.

	\bibitem{JSDM}
	A.~Adhikary, J.~Nam, J.-Y. Ahn, and G.~Caire, ``Joint spatial division and
	multiplexing the large-scale array regime,'' \emph{{IEEE} Trans. Inf.
		Theory}, vol.~59, no.~10, pp. 6441--6463, Oct. 2013.
	
	\bibitem{TQF}
	D.~Kim, G.~Lee, and Y.~Sung, ``Two-stage beamformer design for massive {MIMO}
	downlink by trace quotient formulation,'' \emph{{IEEE} Trans. Commun.},
	vol.~63, no.~6, pp. 2200--2211, Jun. 2015.
	
	\bibitem{iterative}
	J.~Chen and V.~K.~N. Lau, ``Two-tier precoding for {FDD} multi-cell massive
	{MIMO} time-varying interference networks,'' \emph{{IEEE} J. Sel. Areas
		Commun.}, vol.~32, no.~6, pp. 1230--1238, Jun. 2014.
	
	\bibitem{beam_division}
	C.~Sun, X.~Gao, S.~Jin, M.~Matthaiou, Z.~Ding, and C.~Xiao, ``Beam division
	multiple access transmission for massive {MIMO} communications,''
	\emph{{IEEE} Trans. Commun.}, vol.~63, no.~6, pp. 2170--2184, Jun. 2015.
	

\bibitem{sector} { Scanferla, D., ``Studies on 6-sector-site deployment in downlink LTE'', Eindhoven University of Technology, Jan. 2012.}

	
\bibitem{Association}
 \textcolor {blue}{T. Van Chien, E. Bj\"{o}rnson and E. G. Larsson, ``Joint power allocation and user association
optimization for massive MIMO Systems,''
\emph{IEEE Trans. Wireless
Commun.,} vol. 15, no. 9, pp. 6384-6399, September 2016.}

	
	\bibitem{onering}
	A.~Abdi and M.~Kaveh, ``A space-time correlation model for multielement antenna
	systems in mobile fading channels,'' \emph{{IEEE} J. Sel. Areas Commun.},
	vol.~20, no.~3, pp. 550--560, Apr. 2002.
	

	
	\bibitem{gao_E}
H. Xie, F. Gao and S. Jin, ``An Overview of Low-Rank Channel Estimation for
Massive MIMO Systems,'' \emph{IEEE Access}, vol.~4, no., pp.~7313-7321, Nov. 2016.


%
%
	
	
\end{thebibliography}
%

\end{document}